\documentstyle{article}

\author{Zhen Wang\\
Physics Department, LiaoNing Normal University, DaLian, PC:116029 P. R. China}
\title{The Time's Arrow within the Uncertainty Quantum
}
\input tcilatex

\QQQ{Description}{
}

\QQQ{Subject}{
}

\QQQ{Keywords}{
uncertainty, quantum, time's arrow
}

\QQQ{Typist}{
}

\QQQ{Created}{
}

\QQQ{Last Revised}{
}

\QQQ{Language}{
American English
}

\begin{document}

\maketitle
\begin{abstract}
A generalized framework is developed which uses a set description instead of
wavefunction to emphasize the role of the observer. Such a framework is
found to be very effective in the study of the measurement problem and
time's arrow. Measurement in classical and quantum theory is given a unified
treatment. With the introduction of the concept of uncertainty quantum which
is the basic unit of measurement, we show that the time's arrow within the
uncertainty quantum is just opposite to the time's arrow in the observable
reality. A special constant is discussed which explains our sensation of
time and provides a permanent substrate for all change. It is shown that the
whole spacetime connects together in a delicate structure.
\end{abstract}

\section{Introduction}

In searching for a linkage between the two realms of classical and quantum
physics, many researchers have restricted themselves with highly idealized
models amenable to exact solutions. While being helpful to our knowledge of
the border-crossing between the two realms[1], this may also get too much of
our attention to the well-defined process[2-4], so that less notice is taken
of the simple yet profound similarities of some physical process between the
two realms. That may be partly the reason that some physicists suggest
satirically that papers in theoretical physics nowadays refer only to other
papers (and quite often, only to other papers in theoretical physics). To
our knowledge, very few people have study the implication of the quantum
aspect of measuring process in classical physics. By that I mean, while so
much unfruitful effort has been made to understand the meaning of quantum
measurement, can we bypass the difficulty by looking for masked quantum
behavior in classical area? A way for doing this is to examine the
similarity in measuring process between classical and quantum physics. We
know an observer is needed in order for the wavefunction to collapse (from
?) to reality, while in classical physics the observer seems unnecessary. In
fact, in both cases a measuring unit is needed for the measuring result to
be meaningful, which has the quantum feature, i.e. subjectively uncertainty.
Measuring unit and the result together compose a measurement. In both
classical and quantum physics, the Second Law is observed for all
measurement results. Does it have any difference for the measuring unit?
Directed under such notion, we find some interesting and profound aspects
for time's arrow, as well as a generalized framework to reconcile classical
and quantum physics.

Time's arrow has been one of the most puzzling problem in modern physics.
Though inner-time scale[5] in various phenomenological problems adds
complementally to the second law, no satisfactory explanation out of those
precise and reversible fundamental physics theory has been found. What has
been made clear is, it seems to me, the problem is related to other
difficulties just as great. For examples, Poincare Recurrence[6] hints that
time's arrow may be related to the meaning of infinite in mathematics. And
Schroedinger Cat and other paradoxes[7] in quantum theory show, according to
some physicists, that measuring process is time-unsymmetrical so that time
seems related in a complex way to a more mind-boggling concept, spirit. It's
a wonderful idea to consider[8] that moments separated in time, just like
elements separated in space, are generally ''non-causally and non-locally
related projection'' of a higher-dimensional reality. But I never think that
a description with higher dimension will lead to a better understanding of
the problem, simply because it's only mathematical skill, not true in
reality. Physics should lead to, I believe, an overall and self-contained
understanding of the world, not just empirical skills in prediction.

Therefore much more revolutionary changes in ideology is needed to
understand the meaning of time. This may account for the present difficulty
in the research of time and related problems. In developing the widely
expected theory of quantum gravity, some physicists[7] believe that time is
deeply involved in the theory, and even suggest that the success of the
theory depends on some great changes to our idea of space and time. I quite
agree on this point. But I would like to argue in this paper that isn't the
problem caused by our mixing up the two different concepts, time in physics
and time in reality? As has been well established since Boltzmann, time's
arrow in physics is generally referred to an irreversible process, which is
unsymmetrical in time. We have been so sure of this because reversible
process gives no direction, yet time {\bf in reality} does exist. Here I'd
like to point out that it may be a big mistake to relate time with
irreversibility, which has deep repercussion. Time embodies itself not only
in irreversible changes, but in all changes including reversible changes.
With this notion, we can naturally develop an understanding of time by
examining the basic feature of measurement and the similarity in measuring
process between classical and quantum physics. We shall see that quantum
theory, with some extension to what has been expressed with ''multi-world''
theory[9], can give self-contained explanation for time without resort to
the paradoxical anthropic principle.

Therefore we are going to show that time and the measuring difficulty in
quantum physics can be considered as different projections of a single
problem. Comparing with quantum physics, we are going to point out that all
measurable quantities in classical physics, especially space, time and mass,
are in quanta (uncertainty quanta, as we call it), which are the smallest
uncertainty units that compose the observable quantities. By designing a
generalized framework of observables, we can describe a system with a set of
possible states. This is different from the description with a wave function
which only depicts a virtual reality, to be ready to collapse with
measurement. We shall show in later works that in such a generalized
framework, causality restriction established by relativity and nonlocal
correlation in quantum theory can be reconciled. When this new method of
description is applied to the whole universe, we find a simple yet profound
symmetry for all possible states of the universe. More important, this
symmetry is employed to show that time's arrow within the uncertainty quanta
is just opposite to the time's arrow in our observable reality. It is also
possible to see the nature of irreversibility within such a framework. All
these conclusions support the philosophical viewpoint of quantum theory that
any relation is meaningless unless we clearly indicate the observer or
subject.

\section{State Set and Time's Arrow}

In quantum physics, the state of a system is described with a wave function,
whose evolution is dominated by a deterministic equation. The most important
thing here is, the wave function is not the reality. Reality is the result
of measurement of some observables, whose probabilities can be calculated
from the wave function, the {\bf virtual reality}. Thus in this precise
theory reality has no real meaning without measurement. This is the basic
difference between classical and quantum physics, and also the first point
where we are to make some discussion. In our research, we try to deal
directly with reality, rather than the virtual reality described with wave
function. That is, we concentrate on what variables describe the system,
rather than what the wavefunction will evolve. There is no point to talk
about the state of a system without measurement. In doing so, the role of
the observer is emphasized. So we define a set of states of some
observables, the state set, to describe the present state of the system in
research. The state of the system in the next moment may be described with
another state set. Thus the evolution of a system is described with a series
of state sets in different moments, which, with the name of uncertainty
quanta of time, naturally depict a picture of quantized time. Here we shall
not talk about the dynamical law dominating the evolution of the state set
in time, because we are studying time itself. Rather, from the perspective
of a special symmetry between different state sets, we shall see the nature
of time. We hope that such a set description will still be fruitful in the
research of other problems.

The second important point in our research is that we always concentrate on
the role of the observer. We shall never discuss an absolute object without
any observer. Any system in research is regarded as the environment of a
certain observer. When the system is all of our research, it is all of our
environment. Thus the state set actually describes the environment or the
state of an observer. That is, we concentrate on the observer, rather than
an objective system which is always the environment of an observer. We put
special stress on this point because it seems to me that the measuring
difficulty in quantum theory arises mainly from the absolute objectivity of
reality that has been widely admitted since Galileo's time. As will be
shown, such absolute objectivity is just as questionable in classical
physics as in quantum theory.

In order to introduce some important concepts, a generalized framework of
quantum theory is needed. We divide all observables into three groups: the
present observables, the outer observables, and the inner observable. The
present observables describe the state of the environment (e.g. the system
in our research) at the present moment. The inner observables are those that
are not commutable with the present observables. All possible states (or
values) of inner observables compose the inner environment (referred as IE
in the following). The outer observables are those that are commutable with
the present observables. All possible states of outer observables compose
the outer environment (referred as OE in the following). The state set
comprises the measured state of the present observables and outer
environment. It is evident that OE can be observed at present state while IE
can not. Another important difference between IE and OE is: all states in OE
are equivalent while all states in IE are totally uncertain. This is because
that all states in OE are measurable but we do not measure, or do not
distinguish them. Therefore these states are equivalent to us observer and
do not give us the sense of time. That is, OE gives no sense of time and all
states in OE are time symmetric. But we can talk nothing about IE because we
have not a bit of information about it. It may have any unexpected amount,
structure or property. As we shall discuss in other work, this complete
uncertainty of IE makes convergence in many cases precarious.

Now we are ready to define the most important observable, $A_0$, which is
defined to be commutable with all inner and outer observables. Since we give
no restrictions to the symmetry of the variables, $A_0$ can be nothing but
only a constant. That is, its eigenvalue is a constant, never changing in
time. As we discussed in the introduction of this paper, time is related
with change in our theory. Therefore we can not talk about time before we
understand the meaning of change. This requires that we have something that
never changes. All dynamics fix the meaning of time to study the meaning of
change. This has to be reversed if we are to study the meaning of time. The
constant $A_0$ means we now have found something perpetual. This is
important because it provides a perpetual substrate for all changes. Without
it, change would have no meaning, so that we have no ways to discuss time.
It is this constant that enables us to sense the time. Since $A_0$ is
commutable with all observables and has only one possible value, it is
always a present observable. When $A_0$ is the only present observable, all
states of the environment are in the OE and the IE is null. In such a case
OE is the Whole Environment (referred as {\it WE} in the following). We
express {\it WE} as a set composed of all possible states of the world, in
which all states are equivalent or time-symmetric because $WE$ is the OE of $%
A_0$ .

$$
WE=\{\phi _i\mid P\phi _i=\phi
_i,\;for\;all\;\;i\}\;\;\;\;\;\;\;\;\;\;\;\;\;\;\;\;\;\;\;\;\;\;\;\;\;\;\;\;%
\;\;\;\;\;\;\;\;\;\;\;\;\;\;\;\;\;\;\;(1) 
$$
where $P$ is the exchange operator related with the symmetry in OE. Since $%
A_0$ is always a present observable, this symmetry is perpetual. This
perpetual symmetry has tremendous philosophical meaning: we make all
measurements on such a substrate state in which all states are equivalent
and give no sense of time, and we observe all changes with comparison to a
perpetual substrate state which contains all states and has no change
because all states are equivalent. That is, change means deviation from the
state $A_0$.

Generally an observer always has some specific observables as his present
observables. Therefore what he observes is only a part of $WE$ so that his
state set is a subset of $WE$ .

$$
A=\{\phi _i\mid P\phi _i=\phi _i,\;\phi _i\in
OE^A\,\}\;\;\;\;\;\;\;\;\;\;\;\;\;\;\;\;\;\;\;\;\;\;\;\;\;\;\;\;\;\;\;\;\;\;%
\;\;\;\;\;\;\;\;\;\;\;\;\;\;\;\;(2) 
$$
where $OE^A$ is the OE of state $A$ . Such a state set describes the state
of the observer within that shortest time interval (time quantum). Since
outer observables are commutable with present observables and all states in
OE are equivalent, the OE actually represents the choices that the observer
can make in the state represented by $A$. Thus if we can find new equations
reversibly connecting states which are not reversible to us before, our OE
is then enriched. It is evident that the more ordered observer should have
more states in his OE, which give no sensation of time but just furnish the
environment with symmetric or reversible states (thus an more ordered
observer has slower watch. We'll come to this point later). That means,
reversibility is possible only between states in OE. The states in IE
represent the defect or limitation of the observer system, because they can
not be measured {\em without losing present information}. They are
unsymmetrical compared with the states in OE. We shall show that it is this
part of {\it WE}, the complementary set of the state set, that really gives
the system the sensation of time.

Since $A_0$ is always a present observable, all states in $WE$ are
symmetric. To keep this perpetual symmetry, the observer must have a state
set, which is exactly the complementary set of the present state set, in
another moment. That is, IE and OE will change their role in order to keep $%
A_0$ a perpetual constant. Or, all states comes together with their
respective complementary states, with one to be the {\bf present} and the
other to be the {\bf future} (for some specific observer, of course. This
sort of relativity has already been made known by Einstein). Thus the broken
symmetry of the states in {\it WE} leads to the unfolding of time, in which
the symmetry in {\it WE} gets realized.

In the above picture, we can find the time's arrow that is opposite to the
Second Law, or to that in our present reality. In fact, time exists for an
observer only because the OE is only part of $WE$, i.e., it is the result of
the broken symmetry in {\it WE}. Therefore a perfect observer that has {\it %
WE} as its OE would have no time. Suppose the outer environment of observer
X experiences a process P that can be described consecutively with states A$%
_1$ , A$_2$ , $\ldots $$\ldots $ A$_k$ . Then the time's arrow for X would
be given by the series

$$
T:\;\;A_1,A_2,\ldots \ldots A_k 
$$
Because of the perpetual symmetry, all the complement of the states will
come true to the system in other moments. Obviously in order to maintain the
perpetual symmetry they will come true in such an order

$$
T^{\prime }:\;A_1^{\prime },A_2^{\prime },\ldots \ldots A_k^{\prime } 
$$
Since A$_1$ and A$_1^{\prime }$ , A$_2$ and A$_2^{\prime }$ , $\ldots $$%
\ldots $ A$_k$and A$_k^{\prime }$ are all complements to each other, a
physical observable that increases in the process T will decrease in the
process T'. Thus theoretically we may expect that the Second Law is reversed
in other time (before or future) with regard to our present state. Just as
infinite in mathematics can not be clearly figured out, {\em we can not see
this without losing present information} (e.g, classification, function or
nature of observables), because they are in IE to our present state. It
should be noted here that there is no restriction that the observer in the
state A must fall into the state A' immediately. He may continue to choose
new environment from the symmetric states within its state set. This may be
called the free will of the observer. Yet From the perpetual symmetry it is
sure that the process T' will come true for the observer. Then he will lose
some his present information, which has been widely admitted to be the
symbol of time's arrow. That is doomed contingency for any observer system.
Therefore time's arrow exists for all observers, which changes its
embodiment in accordance with the state set.

What I try to emphasize in this paper is we don't need the infinitely long
Poincare Recurrence to see the opposite time arrow. It functions within the
uncertainty quanta as we shall show in next Section. Here we briefly discuss
the matter from the perspective of a perfect observer. For each system X,
there will be a system X' which is the conjugate system of X, because a
perfect observer has $WE$ as its OE. Therefore, according to the perpetual
symmetry of {\it WE}, when an observer system takes a state set A from {\it %
WE}, there must be another observer taking the complement set A' to maintain
the symmetry. This means, the two observers are equivalent and symmetric in
position or status. Both are concealed in the IE, or limitation of
environment of the other observer. Most important, they will exchange their
roles in their respective future to get the perpetual symmetry realized for
each observer. Thus when X experiences an event described as A$%
\longrightarrow $A', its conjugate system X' just simultaneously experiences
the opposite procedure A'$\longrightarrow $A. Both system have their own
second law and time's arrow. But their limitations restrict them from
knowing the opposite process and fail to see the perpetual symmetry. In
fact, both system compose the limitation of the other system. This is just
what happens within the uncertainty quantum, which is the smallest unit of
measurement for an observer. Therefore when entropy increases in our
environment, it decreases within our limitation of measurement, or in the
inner environment that we can not identify because of our limitation.

\section{Measurement and Uncertainty Quantum}

With the change from wave function description to state set description, the
role of measurement has been emphasized. Any state of a system is closely
related to measurement, which is in turn attributed to an observer, thus we
talk about the state of the observer instead of its object system. The
consideration of measurement naturally divides all states into OE and IE.
The collapse of the wave function means nothing but the selection of a new
state set. The definition of observer in physics is just the ability of
selecting state sets. As a matter of fact, life has nothing more than the
ability of distinguishing or selection. Such ability arises because an
observer always has $A_0$ as his present variable. Thus wavefuction
collapses because we try to analyzes the state of other moment with present
classification of states.

When the new state set is a subset of the old one, present information may
be kept. But if it is not a subset of the old state set, the measurement 
{\em can not be done without losing information}. We put special emphasize
on this point to avoid misunderstanding. Therefore a measurement may be a
border-crossing between OE and IE, or it may be a symmetry-breaking in OE.
Such clarity for subject is especially important in the research of the
foundation of physics because it is always impossible to identify an object
to such an absolute clarity that is independent of any observer, i.e.,
measurement always has its limit. This is {\bf not} a trivial technical
problem. Yet far less than enough attention has been paid to it even since
quantum theory was widely accepted. It is usually considered that the key to
the problem of measurement in quantum theory may hide in the further
understanding of quantum theory. But actually it may also hide in the
further understanding of measurement. Without measurement, the concept of
environment can not hold so that observer has no meaning.

In order to measure, we need to have a unit. In some problems, the result of
measurement does not change with the adopted unit. But in problems as in
fractal geometry [10], it does. Anyway there must be a smallest unit that
can not be measured itself. This is the {\bf essence} of measurement. This
is obviously true for all observable quantities, since no quantity can be
made up of zeros. We call this smallest unit uncertainty quantum. It is this
uncertainty quantum that endows a quantity with real meaning. Here we would
like to discuss a thought experiment to show that time is in quantum.

Suppose we have an ideal, infinitely-high-rate camera which can take
infinite films in any short time interval. Of course we also have an ideal
projector. Then let's aim at a running dog. If time were not quantized, we
would be able to take infinite films of the running dog in any time
interval. If then we show these infinite number of films at the normal rate
in cinema, the picture on the screen would be motionless. Of course we can
see a running dog if we show the films at the same rate as we take them. But
because of our persistence of vision (which may be a little bit different
from person to person), we lose infinite information of the reality. We
can't know what happens within the interval of persistence of vision. That
shows motion is the direct result of our persistence of vision, which is the
uncertainty quantum of time for us human observer in this case.

Thus (the outer) environment is the result of measurement of the observer
with the uncertainty quantum to be the unit. Therefore in a sense, the
uncertainty quantum determines what environment the observer identifies. The
most important property of the uncertainty quantum is that all states or
structures within the quantum are totally unknown {\em in present state}.
(Otherwise we would be able to find structures within it and it would not be
the smallest unit.) That means, all structures within the uncertainty
quantum are beyond the recognition of the observer in present state. Since
the environment of an observer is described with a state set, it is obvious
all structures within the uncertainty quantum are not in the state set.
Therefore, the complement of the state set composes the uncertainty quantum,
which then endows the measurement with real meaning. That is, the IE
actually works as the basic measuring units, the uncertainty quantum, for OE
to be identified. Now that we have linked together OE and IE with the
uncertainty quantum, it can be seen clearly that the opposite time arrow we
discussed above is just within the uncertainty quantum, i.e. within the
limitation of the observer. Therefore non-local and non-casual events
separated in time are connected together by responsible measurement and
uncertainty quantum. All states that seem permanently lost in the
many-worlds quantum theory are now used not only in measurement as the basic
unit, but also in the unfurling of time.

It is also interesting to point out that combining Einstein's theory of
relativity with our idea of uncertainty quantum, we can also arrive at the
conclusion of the opposite arrow of time. In Einstein's theory, no signal
can travel faster than light. Otherwise the arrow of time would be reversed.
The related paradox is: if an observer could travel at a super light speed,
he might be able to meet and kill himself in the past in his childhood.
Though super light speed is permitted in our theory, such a paradox can not
hold in our theory. If $q_l$ is the uncertainty quantum of length (space)
and $q_t$ the uncertainty quantum of time, we may express the speed of light
with

$$
\;\;\;\;\;\;\;\;\;\;\;\;\;\;\;\;c=\frac{q_l}{q_t}\;\;\;\;\;\;\;\;\;\;\;\;\;%
\;\;\;\;\;\;\;\;\;\;\;\;\;\;\;\;\;\;\;\;\;\;\;\;\;\;\;\;\;\;\;\;\;\;\;\;(3) 
$$
Like Einstein's theory of relativity, no signals are permitted to travel
faster than light, but the light speed, which is always the maximum speed
for an observer, may change its value and may be different for different
observers. This is because the uncertainty quanta may change. If the
uncertainty quantum of time changes greatly enough, we might have to change
our definition for past, present and future to such a extent that our {\bf %
past} defined in past becomes part of our present. That is, the past,
present and future is only the result of measurement with our {\bf present}
uncertainty quantum of time. The smaller the time quantum, the slower the
watch of the observer. The larger of the speed of light for one observer,
the more non-local correlation to another observer with smaller light speed.
Therefore, time's arrow will never be reversed and there will never be
paradoxes as two selves for one person.

The surface of the light cone divides all spacetime into two parts: the
spacelike part and the timelike part. From (3) we can see that within the
spacelike area, a signal can not travel farther than a space quantum in one
time quantum, while in the timelike area a signal may cross over any space
quantum in one time quantum. In the perspective of our framework, to an
observer with such a lightcone the timelike area corresponds to smaller time
quantum and larger space quantum. (These two are the same, we shall discuss
this elsewhere.) Therefore the spacelike area corresponds to OE in our
theory and timelike area IE. OE may be called causality area, and IE may be
called correlation area. Quantum correlation as in experiments of Bell's
inequality always happens within IE of the observer, where smaller time
quantum is needed to see its structure, implicating a super light speed and
a reversed arrow of time to the observer.

The idea of uncertainty quantum also has its implication in mathematics. In
an {\bf infinite} measuring process, all structures within the uncertainty
quantum are unfolded and realized. Therefore all our limitations in space,
time and mass are contained in the corresponding uncertainty quantum. All
our limitations are {\bf connected}, and time arrow is only one way to
embody the limitation. The uncertainty quantum is another way. This is the
essential ignorance of the observer. Measurement is the process in which a
subset of $WE$ is chosen to be the definite ignorance. We know nothing about
this ignorance, especially, the number of states in IE. This is exactly
where comes the concept {\bf infinite}. In mathematics the single counting
process can be carried on endlessly. This process is just the unfolding of
the points within some basic unit. That's why the infinity of the points in
the whole axis and the infinity of the points in a segment are of the same
order. In such linear process there is no loss of information. But in
nonlinear problems as we shall discuss in later works, we have to face the
breaking of the uncertainty quantum, which always means loss of information
and irreversibility.

\section{Conclusion}

The observer in this research is not necessarily a man. It may be a man with
scientific apparatus as his sense organ. It may even be a theoretical system
complex enough. In such a generalized framework we show that we can get some
new insight into the measurement problem by using a state set description,
which is linked with the basic principle of measurement in both classical
and quantum physics. This description is necessary in revealing the two
opposite directions of the times arrow. We can also understand in this
framework why we only sense one time's arrow. The other arrow of time
amazingly hides deeply within the uncertainty quanta, which compose the
result of our observation. Although by definition the uncertainty quantum
can not be identified, it may have observable effect[11]. We shall show in
our later works that locality and quantum correlation can be in a good
consistence which is more reasonable in philosophy.

{\bf REFERENCES}

1. W. H. Zurek, Phys. Today, {\bf 44}(10), 36(1991)

2. A. O. Caldeira and A. J. Leggett, Phys. Rev., {\bf A31}, 1059(1985)

3. W. G. Unruh and W. H. Zurek, Phys. Rev., {\bf D40}, 1071(1989)

4. G. T. Moore and M. O. Scully, {\it Frontier of Nonequilibrium Statistical
Physics}, edited by W. H. Zurek, (Plenum, New York, 1986)

5. I. Prigogin, {\it From Being to Becoming} (W. H. Freeman and Company,
1980)

6. B. Misra, {\it Proc. Natl. Acd. Sci. U.S.\/}, {\bf 75}, 1629(1978)

7. R. Penrose, {\it The Emperor's New Mind, \/}(Oxford University Press,
1989)

8. D. Bohm, {\it Wholeness and The Implicate Order}, (Routledge \& Kegan
Paul, 1981)

9. B. S. Dewitt \& N. Graham, {\it The Many-Worlds Interpretation of Quantum
Mechanics\/} (Princeton University Press, 1973)

10. B. Mandelbrot, {\it The Fractal Geometry of Nature\/} (New York:
Freeman, 1977)

11. Zhen Wang, quant-ph/9804070

\end{document}